\documentclass[aps,prl,preprint,groupedaddress,notitlepage]{revtex4}
\usepackage{graphicx}

\preprint{NuFact15 - Rio de Janeiro, Brazil - August 2015}

\begin{document}


\title{Results and Prospects from T2K}
\thanks{\it Presented at NuFact15, 10-15 Aug 2015, Rio de Janeiro, 
Brazil [C15-08-10.2]}



\author{Kirsty Duffy}
\email[]{kirsty.duffy@physics.ox.ac.uk}
\thanks{Speaker}
\affiliation{University of Oxford}


\date{\today}

\begin{abstract}
As measurements of the neutrino oscillation parameters improve it is becoming more interesting to study antineutrino oscillations, to investigate CP and CPT violation in the lepton sector and nonstandard matter effects.
We present the most recent T2K antineutrino oscillation results, from data collected using a $\overline{\nu}_{\mu}$-enhanced neutrino beam corresponding to $4.01 \times 10^{20}$ protons on target. 
 The first analysis of $\overline{\nu}_e$ appearance at T2K is presented, as well as world-leading measurements of the dominant oscillation parameters for $\overline{\nu}_{\mu}$ disappearance. T2K measures $\sin^2\overline{\theta}_{23} = 0.46^{+0.14}_{-0.06}$ and $\Delta \overline m^2_{32} = 2.50^{+0.3}_{-0.2} \times 10^{-3} \mbox{ eV}^2$, which is consistent with previous T2K measurements of the neutrino oscillation parameters and existing antineutrino measurements.
\end{abstract}


\maketitle


\section{The T2K experiment\label{sec:t2k}}

T2K is a long-baseline neutrino oscillation experiment located in Japan, which uses the 30 GeV proton beam from the J-PARC accelerator to create a muon neutrino beam. The proton beam is directed onto a graphite target, and the resulting pions focussed by magnetic horns (which can select either $\pi^+$, for a beam composed mainly of $\nu_{\mu}$, or $\pi^-$, for a beam composed mainly of $\overline{\nu}_{\mu}$), into a 96 m decay tunnel. The neutrino beam is measured by two near detectors located 280 m from the target, and a far detector, Super-Kamiokande. 
The far detector and one of the near detectors are placed 2.5$^{\circ}$ off-axis with respect to the neutrino beam, which results in a quasi-monochromatic neutrino energy spectrum that is sharply peaked around 0.6 GeV. The baseline between neutrino production and the far detector, 295 km, is carefully chosen to correspond to the first minimum in the $\nu_{\mu}$ survival probability at the peak energy. T2K can measure neutrino oscillation in two channels: $\nu_{\mu}$ disappearance (which is dominated by the oscillation parameters $\sin^2\theta_{23}$  and $\Delta m^2_{32}$) and $\nu_e$ appearance (which is sensitive to $\sin^2\theta_{13}$ and $\delta_{CP}$).


T2K has been taking data since 2010, and has so far collected $1.1\times 10^{21}$ protons on target (POT). The beam power has been steadily increasing, and stable running at 345 kW was achieved in 2015, with a maximum beam power of 371 kW.
 Since mid-2013 the beam has been running in antineutrino mode, where the horn currents are reversed to select $\pi^-$ instead of $\pi^+$, resulting in a beam that is mostly composed of antineutrinos. Sensitivity studies show that T2K may, depending on the value of $\delta_{CP}$, be sensitive to $\delta_{CP}$ when roughly equal amounts of neutrino-mode and antineutrino-mode data are collected, with the full predicted data set of $7.8 \times 10^{21}$ POT. In addition, measurements of antineutrino oscillations will allow us to test the PMNS framework and search for CP violation (if $P(\overline{\nu}_{\mu} \rightarrow \overline{\nu}_e) \neq  P(\nu_{\mu} \rightarrow \nu_e)$) or CPT violation or non-standard matter effects (if $P(\overline{\nu}_{\mu} \rightarrow \overline{\nu}_{\mu}) \neq  P(\nu_{\mu} \rightarrow \nu_{\mu})$). $4.011 \times 10^{20}$ POT have been collected in antineutrino mode, which is roughly one third of the total data set. However, the event rates in antineutrino mode are significantly lower than in the neutrino-mode beam, due to both pion multiplicity and the difference between $\nu$ and $\overline{\nu}$ cross sections.

\subsection{The near detectors: ND280 and INGRID\label{sec:ND280INGRID}}

\begin{figure}[bh]
	\includegraphics[width=0.35\textwidth]{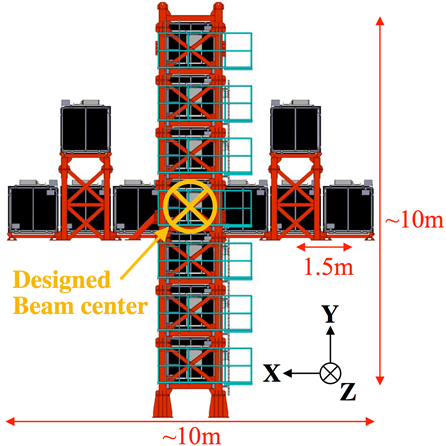} 
	\includegraphics[width=0.45\textwidth]{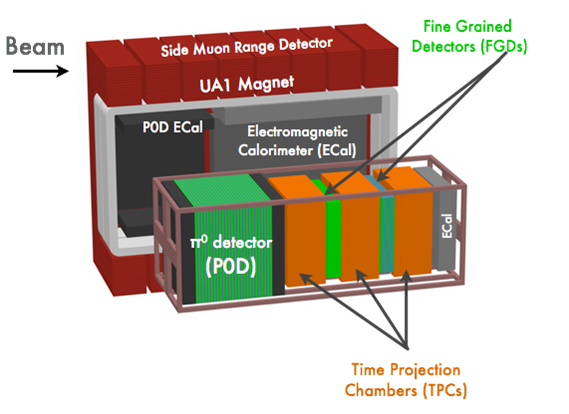} 
	\caption{\label{fig:nds}The T2K near detectors: INGRID (\textbf{left}) and ND280 (\textbf{right}).}
\end{figure}

T2K has two near detectors, shown in figure~\ref{fig:nds}: INGRID, which is on-axis with respect to the neutrino beam, and ND280, which is at the same off-axis angle as Super-Kamiokande. Both detectors have a rich program of physics which has been covered in other presentations at this conference; here we will focus on their use in the oscillation analysis.

The on-axis detector, INGRID, is an array of 7+7 iron/scintillator detectors arranged in a cross shape centered on the beam axis. It is used to measure the beam stability, profile, and direction, and has shown that the beam direction is stable to within 0.4 mrad.

The off-axis detector, ND280, is used directly in the oscillation fits to reduce the flux and cross-section uncertainties. It is made up of many subdetectors, but only the 
central Fine-Grained Detectors (FGDs) and Time Projection Chambers (TPCs) are used for this analysis. ND280 contains two FGDs, which provide a target for neutrino interactions with excellent vertexing capabilities at the interaction point. The current oscillation fits use only FGD1, for which the target material is carbon. The TPCs are then used to measure the interaction products, and give very good momentum resolution and particle identification. ND280 is contained in the repurposed UA1 magnet, which enables the TPC information to distinguish positive and negative charged leptons from $\overline{\nu}$ and $\nu$ interactions.

\subsection{The far detector: Super-Kamiokande\label{sec:SK}}

The far detector, Super-Kamiokande, is a 50kton water Cherenkov detector (22.5kton fiducial mass). 
 It has no magnetic field so cannot distinguish between neutrino and antineutrino interactions, but is capable of very good $\mu$/$e$ separation by the pattern of light from the charged lepton ($<$1\% of $\mu$ events are misidentified).

\section{Oscillation analysis strategy\label{sec:OA}}

The analysis strategy for the oscillation results presented here is similar to previous T2K results~\cite{T2K_longosc}: data samples of charged current (CC) interactions are fit at ND280 to provide a tuned prediction of the unoscillated spectrum at the far detector and its associated uncertainty. This is then compared to the data at the far detector, where $\mu$-like or $e$-like data samples are fit to estimate the oscillation parameters.

\subsection{Near detector (ND280) fit\label{sec:ndfit}}

The near detector fit takes inputs from 
theoretical flux, cross section, and detector models, each of which has its own uncertanties. 

The flux model uses information from measurements by INGRID and the beam monitors, as well as external data from the NA61/SHINE experiment. It is used to constrain the prediction at the far detector through correlations between the neutrino flux at both detectors, as predicted by beam simulations. 

The predictions at both ND280 and Super-Kamiokande use the same cross-section model, so the ND280 fit can reduce the cross-section uncertainty in the Super-Kamiokande prediction by fitting parameter values in the underlying models. Information from external data (from the MINER$\nu$A and MiniBooNE experiments) are also included as a prior for the ND280 fit. It is not possible to constrain all the cross-section parameters because of the different target materials in the near detector (carbon) and far detector (primarily oxygen), as the relative errors between interactions on carbon and oxygen are not always well understood. Separate parameters are used for Fermi momentum, binding energy, multinucleon event normalisation and CC coherent pion production nomalisation on oxygen, which are not well constrained by the near detector. We use a conservative (100\% error, with no correlation with multinucleon events on carbon) ansatz for the multinucleon normalisation on oxygen.

The near detector fit also estimates correlations between the flux and cross-section parameters at Super-Kamiokande.

The data in the near detector are fit in the momentum and angle of the outgoing lepton from the neutrino interaction. Data from the neutrino-mode beam as well as antineutrino mode are used to ensure that the model parameters are consistent between neutrinos and antineutrinos, and provide a constraint on the wrong-sign background ($\nu$ in the $\overline{\nu}$ beam). In total $5.82 \times 10^{20}$ POT of data in neutrino mode and $0.43 \times 10^{20}$ POT of data in antineutrino mode are used, split into seven samples. The antineutrino-mode data are split into `$\nu_{\mu}$ CC 1 track', `$\nu_{\mu}$ CC $>$1 track', `$\overline{\nu}_{\mu}$ CC 1 track', and `$\overline{\nu}_{\mu}$ CC $>$1 track' samples. The `1 track' samples are dominated by CC quasielastic (CCQE) interactions ($\nu_{\mu}+n \rightarrow \mu^- + p$ or $\overline{\nu}_{\mu} + p \rightarrow \mu^+ + n$, the `signal' at Super-Kamiokande), and the `$>$1 track' samples are used to estimate the background from other interactions. The neutrino-mode data are divided into three subsamples according to the number of measured pions associated with the interaction: `$\nu_{\mu} \mbox{ CC}0\pi$', `$\nu_{\mu} \mbox{ CC}1\pi^+$', and `$\nu_{\mu}$ CC other', which are dominated by CCQE, CC resonant pion production, and deep inelastic scattering interactions respectively. 

\begin{figure}[h]
  \begin{tabular}{p{0.5\textwidth}p{0.5\textwidth}}
	\vspace{7pt} \includegraphics[width=0.49\textwidth, height=6.2cm]{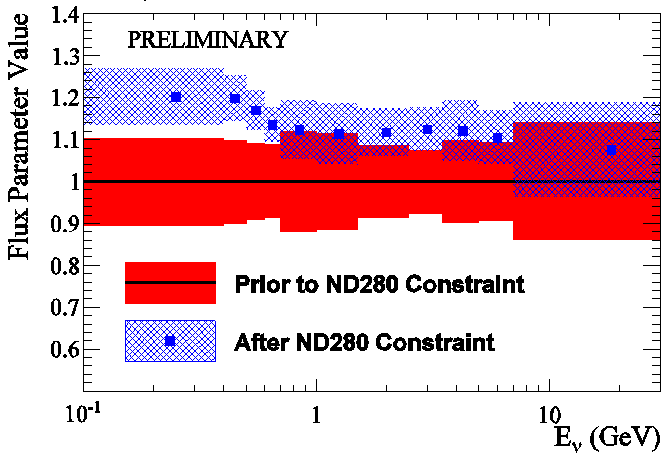} &
	\vspace{0pt} \includegraphics[width=0.49\textwidth]{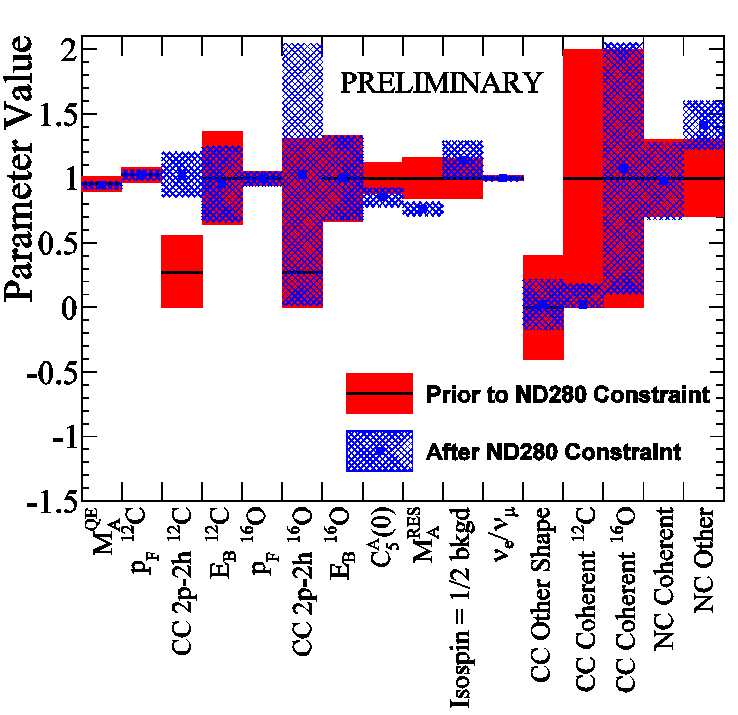}
  \end{tabular}
	\caption{\label{fig:ndfit}Some flux and cross-section systematic parameters with uncertainties before and after the near detector fit. \textbf{Left:} flux parameters for $\overline{\nu}_{\mu}$ flux in the antineutrino-mode beam. \textbf{Right:} underlying parameters for the cross-section models.}
\end{figure}

Figure~\ref{fig:ndfit} shows some of the flux and cross-section parameters with their associated uncertainties before and after the near detector fit. The predicted flux at Super-Kamiokande is generally increased by the fit, although the uncertainty is decreased. Some of the cross-section parameters fit to values which are significantly different to their prior predictions, particularly the multinucleon event normalisation parameter on carbon (``CC 2p-2h $^{12}$C"). The uncertainties on the parameters which the near detector is sensitive to are generally decreased in the fit, but there is not much change to the uncertainties of the oxygen-specific parameters. 

Table~\ref{tab:SKerr} shows the uncertainty in the predicted number of $\overline{\nu}_{\mu}$ events at Super-Kamiokande due to each source of systematic error. The near detector fit reduces the error due to the parameters that it can constrain from 9.2\% to 3.4\%. However, the overall error is dominated by the cross-section parameters which are not constrained by the near detector (in particular the multinucleon event normalisation systematic, which alone produces a 9.5\% uncertainty).

\begin{table}
	\caption{\label{tab:SKerr}Fractional error in the number of $\overline{\nu}_{\mu}$ events predicted at Super-Kamiokande (SK) due to different sources of systematic error, before and after the near detector (ND) fit.}
	\begin{ruledtabular}\begin{tabular}{c|cc}
	    Systematic & $\frac{\Delta N_{SK}}{N_{SK}}$ without ND & $\frac{\Delta N_{SK}}{N_{SK}}$ with ND \\
            \hline
            All common to ND/SK & 9.2\% & 3.4\% \\
            Multinucleon effect on oxygen & \multicolumn{2}{c}{9.5\%}\\
            All oxygen cross-section & \multicolumn{2}{c}{10.0\%} \\
            Final state interaction/secondary interaction at SK & \multicolumn{2}{c}{2.1\%} \\
            SK detector & \multicolumn{2}{c}{3.8\%} \\
            \hline
            Total & 14.4\% & 11.6\% \\
	\end{tabular}\end{ruledtabular}
\end{table}

\subsection{Far detector fit}

The result of the near detector fit is propagated to the far detector. This is used as a prior for the far detector fit, which includes additional uncertainties from a Super-Kamiokande detector model. Two different analyses are described in the following sections, which both use the same near detector fit results but different assumptions and data samples at the far detector.

\section{New results from antineutrino running\label{sec:antinu}}

\subsection{$\overline{\nu}_e$ appearance analysis\label{sec:nuebarapp}}

The aim of this analysis is to look for anti-electron neutrino appearance, separately from electron neutrino appearance (which has already been observed with a significance of 7.3$\sigma$ at T2K~\cite{T2K_nueapp}). In order to test whether or not our data indicate the presence of $\overline{\nu}_e$ appearance, we introduce a new parameter $\beta$ which modifies the $\overline{\nu}_e$ appearance probability:

\begin{equation}
P(\overline{\nu}_{\mu} \rightarrow \overline{\nu}_e) = \beta \times P_{PMNS}(\overline{\nu}_{\mu} \rightarrow \overline{\nu}_e) \nonumber
\end{equation}

\noindent Aside from this, CPT symmetry is assumed (so the same oscillation parameters are used for neutrino and antineutrino oscillations). In this parameterisation, $\beta = 1$ corresponds to $\overline{\nu}_e$ appearance in accordance with the PMNS prediction (which allows for CP violation if $\delta_{CP} \neq 0$). $\beta=0$ corresponds to no $\overline{\nu}_e$ appearance.

The analysis uses a marginal likelihood, which is integrated over all parameters other than $\beta$:

\begin{equation}
\mathcal{L}(\beta) = \int \int \prod_{SK bins} 
\mathcal{L}_{Poisson, bin}(\beta, \vec{o}, \vec{f}) \times \pi_{syst.}(\vec{f}) \times \pi_{osc.}(\vec{o}) d\vec{o} d\vec{f} \nonumber
\end{equation}

\noindent where $\vec{o}$ are the oscillation parameters, $\vec{f}$ are the systematic parameters, $\pi_{syst.}(\vec{f})$ is the prior probability density for the systematic parameters (taken from the near detector fit), and $\pi_{osc.}(\vec{o})$ is the prior probability density for the oscillation parameters. $\mathcal{L}_{Poisson, bin}$ is the Poisson likelihood in each bin given the number of data events and number of predicted events at Super-Kamiokande. The Super-Kamiokande data and prediction are binned in either reconstructed (anti-)neutrino energy ($E_{rec}$) or momentum and angle (with respect to the incoming neutrino direction) of the measured lepton ($p-\theta$), and the product runs over all bins. The oscillation priors are taken from the posterior of the T2K joint $\nu_{\mu}$ and $\nu_e$ fit~\cite{T2K_longosc}, which have a peak value for $\delta_{CP} \sim -\pi/2$.

We report the significance for $\beta = 1$ in two ways: a p-value and a Bayes factor. 

The p-value relies on a test statistic $-2\Delta \ln \mathcal{L} = -2(\ln\mathcal{L}(\beta = 1) - \ln\mathcal{L}(\beta = 0))$, which compares the marginal likelihoods from two fits assuming $\beta = 1$ and $\beta = 0$. This is then compared to the same test statistic calculated from an ensemble of test experiments on fake data generated with $\beta=0$, to characterise how anomalous our data are with respect to the $\beta = 0$ hypothesis.

The Bayes factor is simply the likelihood ratio:

\begin{equation}
B_{10} = \frac{\mathcal{L}( Data | \beta = 1)}{\mathcal{L}(Data | \beta = 0)} \nonumber
\end{equation}

\noindent and describes how much our data favours $\beta = 1$ over $\beta = 0$.

The current data set contains 3 events in the $e$-like sample at Super-Kamiokande. At the T2K best-fit parameter values from neutrino analyses (given in table~\ref{tab:numubar:oscpars}) we expect $\sim$ 1.3 events if $\beta = 0$ and $\sim$ 3.7 events if $\beta = 1$. Examining the full range $\delta_{CP} = [-\frac{\pi}{2},\frac{\pi}{2}]$ and both mass hierarchies, we predict between 3.7 and 5.3 events for $\beta = 1$. Events are selected according to the following criteria: they must be fully contained within the fiducial volume of the detector, have only one reconstructed ring, and have electron-like particle identification. The lepton momentum must be greater than 100 MeV, and the reconstructed (anti-)neutrino energy smaller than 1.25 GeV. The event must pass the $\pi^0$ rejection cuts used in the Super-Kamiokande software and there can be no decay electrons.

Table~\ref{tab:nuebarresults} shows the test statistic ($-2\Delta \ln \mathcal{L}$), the p-value, and the Bayes factor for the data fit. The p-value is greater than 15\% for both the $E_{rec}$ and $p-\theta$ fit, which shows no disagreement between the data and the $\beta=0$ hypothesis. The Bayes factor is 1.1 for the $E_{rec}$ fit and 0.6 for the $p-\theta$ fit (which is equivalent to a Bayes factor of $\sim 1.7$ in favour of $\beta = 0$ over $\beta = 1$). Neither of these show strong enough evidence to support $\beta=1$ over $\beta = 0$, so with the current data set we cannot conclude that we have observed $\overline{\nu}_e$ appearance with statistical significance.

\begin{table}
	\caption{\label{tab:nuebarresults}Test statistics, p-values, and Bayes factors from T2K $\overline{\nu}_e$ appearance analysis}
	\begin{ruledtabular}\begin{tabular}{cccc}
	    Super-Kamiokande binning & $-2\Delta \ln \mathcal{L}$ & p-value & B$_{10}$ \\
	    \hline
	    $\overline{\nu} E_{rec}$ & 0.16 & 0.16 & 1.1 \\
	    Lepton $p-\theta$ & -1.16 & 0.34 & 0.6 \\
	\end{tabular}\end{ruledtabular}
\end{table}


\subsection{$\overline{\nu}_{\mu}$ disappearance analysis\label{sec:numubardisapp}}

This analysis uses the antineutrino-mode data to measure the antineutrino oscillation parameters, so CPT invariance is not assumed. We fit the oscillation parameters that dominate $\overline{\nu}_{\mu}$ disappearance, $\sin^2\overline{\theta}_{23}$ and $\Delta \overline m^2_{32}$, in the ranges given in table~\ref{tab:numubar:oscpars}. All other antineutrino and neutrino oscillation parameters are fixed such that ${\theta}_{12} = \overline{\theta}_{12}$, ${\theta}_{13} = \overline{\theta}_{13}$, $\Delta m^2_{12} = \overline\Delta m^2_{12}$, and $\delta_{CP} = \overline{\delta}_{CP}$.

\begin{table}
	\caption{\label{tab:numubar:oscpars}Oscillation parameters used for the $\overline{\nu}_{\mu}$ disappearance analysis. Fixed values are taken from the output of the T2K $\nu_{\mu} + \nu_e$ joint fit~\cite{T2K_longosc} and the 2014 edition of the Particle Data Booklet~\cite{PDG2014}.}
	\begin{ruledtabular}
		\begin{tabular}{ccc}
		 & Value for neutrinos & Value for antineutrinos \\
		 \hline
		 $\sin^2\theta_{23}$ & 0.527 & 0--1\\
		 $\Delta m^2_{32} (\times 10^{-3} \mbox{eV}^2)$ & 2.51 & 0--20 \\
		 $\sin^2\theta_{13}$ & \multicolumn{2}{c}{0.0248} \\
		 $\delta_{CP}$ (rad.) & \multicolumn{2}{c}{-1.55} \\
		 $\sin^2\theta_{12}$ & \multicolumn{2}{c}{0.304} \\
		 $\Delta m^2_{21} (\times 10^{-5} \mbox{eV}^2)$ & \multicolumn{2}{c}{7.53} \\
		\end{tabular}
	\end{ruledtabular}
\end{table}

The Super-Kamiokande fit maximises a marginal likelihood $\mathcal{L}$ with respect to $\sin^2\overline{\theta}_{23}$ and $\Delta \overline m^2_{32}$:

\begin{equation}
\mathcal{L}(\vec{o}) = \int \prod_{SK bins} \mathcal{L}_{Poisson, bin}(\vec{o}, \vec{f}) \times \pi_{syst.}(\vec{f}) d\vec{f}
\end{equation}

\noindent where all symbols are as previously defined. In this analysis the Super-Kamiokande data was binned only in reconstructed (anti-)neutrino energy.

There are 34 events in the $\mu$-like sample from antineutrino beam data. The event selection is similar to that used in the $e$-like selection: events must be fully contained in the fiducial volume and may only have one reconstructed ring. However, in this case the event must be consistent with a muon-like particle identification, there must be one or fewer decay electrons, and the outgoing lepton momentum must be greater than 200 MeV.

\begin{figure}[h]
	\includegraphics[width=0.49\textwidth]{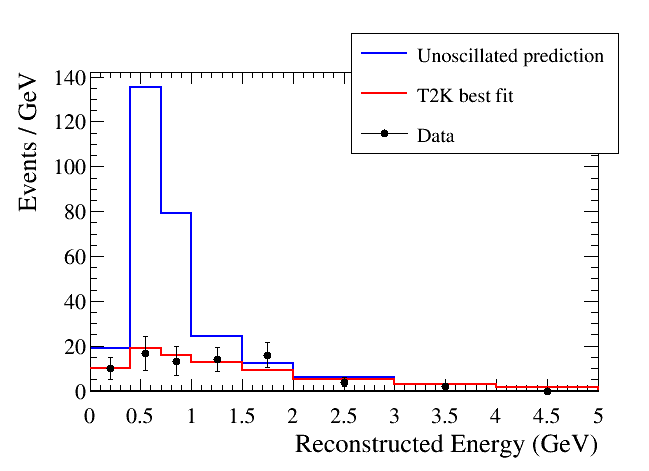}
	\includegraphics[width=0.49\textwidth]{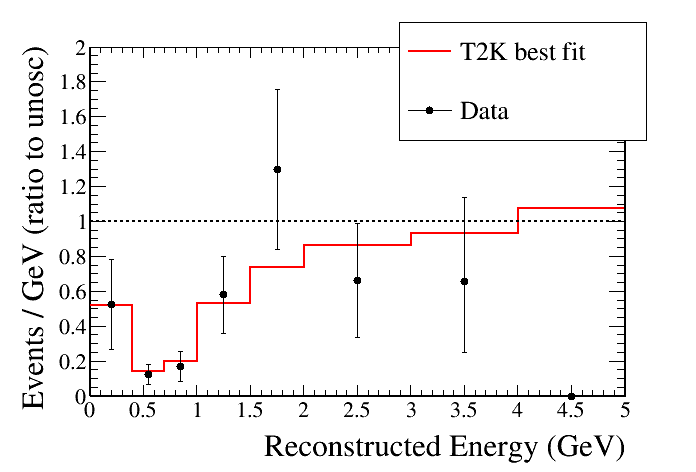}
	\caption{\label{fig:numubar:erec}\textbf{Left:} Predicted reconstructed energy spectrum in the case of no oscillations, and best-fit reconstructed energy spectrum after the data fit. Data is overlaid in black, with statistical errors shown. \textbf{Right:} Ratio of best-fit energy spectrum and data to prediction without oscillations.}
\end{figure}

Figure~\ref{fig:numubar:erec} shows the predicted reconstructed energy spectrum under the no-oscillation hypothesis, and the best-fit spectrum from the data fit, along with the data. The right-hand plot shows the ratio of both data and best-fit spectrum to the unoscillated prediction, which shows the characteristic `oscillation dip' that is clear evidence of $\overline{\nu}_{\mu}$ oscillation. The binning shown in the plot is coarser than the binning used for the data fit.

\begin{figure}[h]
	\includegraphics[width=0.49\textwidth]{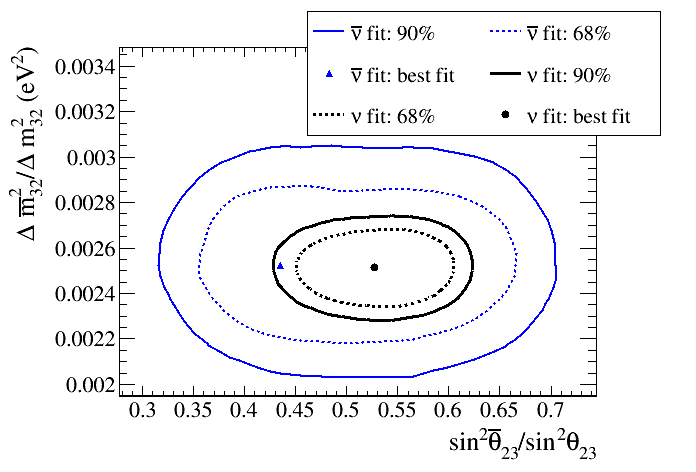}
	\includegraphics[width=0.49\textwidth]{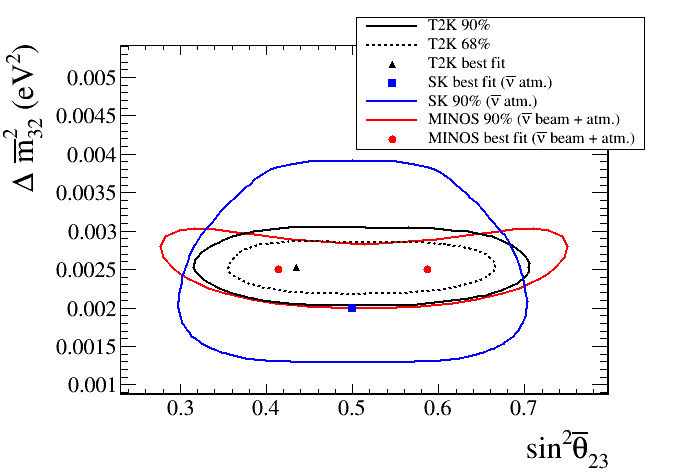}
	\caption{\label{fig:numubar:contours}68\% and 90\% credible intervals in $\sin^2\overline{\theta}_{23}$--$\Delta \overline m^2_{32}$ from the $\overline{\nu}_{\mu}$ disappearance analysis. \textbf{Left:} overlaid with contours from the T2K $\nu_{\mu} + \nu_e$ joint fit~\cite{T2K_longosc}. \textbf{Right:} overlaid with contours from similar analyses by MINOS (using antineutrino-mode beam and atmospheric data)~\cite{MINOS_numubar} and Super-Kamiokande (using atmospheric data only)~\cite{SK_numubar}. The MINOS contour was originally presented in terms of $\sin^22\overline{\theta}_{23}$ and had to be unfolded (hence the two best-fit points).}
\end{figure}

The left hand side of figure~\ref{fig:numubar:contours} shows the 68\% and 90\% credible interval contours in $\sin^2\overline{\theta}_{23}$--$\Delta \overline m^2_{32}$ compared to the contours from the T2K $\nu_{\mu} + \nu_e$ joint fit~\cite{T2K_longosc}. The antineutrino analysis has much larger contours because it has much lower statistics than the neutrino analysis, but the results are consistent: we see no evidence for CPT violation. Projecting the posterior onto one dimension gives best-fit estimates with 1$\sigma$ errors:

\begin{displaymath}
\sin^2\overline{\theta}_{23} = 0.46^{+0.14}_{-0.06} 
\end{displaymath}
\begin{displaymath}
\Delta \overline{m}^2_{32} = 2.50^{+0.3}_{-0.2}\times 10^{-3} \mbox{ eV}^2 
\end{displaymath}

The right hand side of figure~\ref{fig:numubar:contours} shows the contours from this analysis overlaid with the 90\% contours from $\overline{\nu}_{\mu}$ disappearance analyses in MINOS~\cite{MINOS_numubar} and Super-Kamiokande using atmospheric neutrinos~\cite{SK_numubar}. 
The results from all three experiments are in agreement.




\section{Summary and future prospects}

We have presented here the first T2K results based on antineutrino data, including analyses of $\overline{\nu}_e$ appearance and $\overline{\nu}_{\mu}$ disappearance.

For $\overline{\nu}_e$ appearance we calculate a p-value greater than 15\% and a Bayes factor $\sim$ 1, so there is not sufficient evidence to claim observation of $\overline{\nu}_e$ appearance in the current data set.

The 1D best-fit values from the measurement of $\overline{\nu}_{\mu}$ disappearance at T2K are $\sin^22\overline{\theta}_{23} = 0.46^{+0.14}_{-0.06}$ and $\Delta \overline m^2_{32} = 2.50^{+0.3}_{-0.2} \times 10^{-3} \mbox{ eV}^2$. The 2D contours are in agreement with T2K neutrino-mode fits and antineutrino results published by MINOS and Super-Kamiokande.

Both analyses are statistics-limited and T2K is continuing to run with an antineutrino beam which will provide additional data to improve both measurement.



\bibliography{references}

\end{document}